\documentclass[superscriptaddress, twocolumn]{revtex4-1}
\usepackage{amsmath}
\usepackage{graphicx}
\usepackage{inputenc}
\usepackage{color}
\usepackage{setspace}
\usepackage{tabularx}
\usepackage{multirow}
\usepackage{array}
\newcommand{\ndzr}{Nd$_2$Zr$_2$O$_7$}
\newcommand{\hoti}{Ho$_2$Ti$_2$O$_7$}
\newcommand{\tbti}{Tb$_2$Ti$_2$O$_7$}
\newcommand{\erti}{Er$_2$Ti$_2$O$_7$}
\newcommand{\ybti}{Yb$_2$Ti$_2$O$_7$}
\newcommand{\dyti}{Dy$_2$Ti$_2$O$_7$}
\newcommand{\przr}{Pr$_2$Zr$_2$O$_7$}
\newcommand{\nd}{Nd$^{3+}$}

\begin{document}

\author{S. Petit}
\email[]{sylvain.petit@cea.fr}
\affiliation{Laboratoire L\'eon Brillouin, CEA, CNRS, Univerit\'e Paris Saclay, CEA Saclay, F-91191 Gif-sur-Yvette, France}
\author{E. Lhotel}
\email[]{elsa.lhotel@neel.cnrs.fr}
\affiliation{Institut N\'eel, CNRS and Universit\'e Grenoble Alpes, F-38042 Grenoble, France}
\author{B. Canals} 
\affiliation{Institut N\'eel, CNRS and Universit\'e Grenoble Alpes, F-38042 Grenoble, France}
\author{M. Ciomaga Hatnean}
\affiliation{Department of Physics, University of Warwick, Coventry, CV4 7AL, United Kingdom}
\author{J. Ollivier}
\affiliation{Institut Laue Langevin, F-38042 Grenoble, France}
\author{H. Mutka}
\affiliation{Institut Laue Langevin, F-38042 Grenoble, France}
\author{E. Ressouche}
\affiliation{INAC, CEA and Universit\'e Grenoble Alpes, CEA Grenoble, F-38054 Grenoble, France}
\author{A. R. Wildes}
\affiliation{Institut Laue Langevin, F-38042 Grenoble, France}
\author{M. R. Lees}
\affiliation{Department of Physics, University of Warwick, Coventry, CV4 7AL, United Kingdom}
\author{G. Balakrishnan}
\affiliation{Department of Physics, University of Warwick, Coventry, CV4 7AL, United Kingdom}

\title{Observation of magnetic fragmentation in spin ice}

\maketitle

{\bf Fractionalised excitations that emerge from a many body system have revealed rich physics and concepts, from composite fermions in two-dimensional electron systems, revealed through the fractional quantum Hall effect \cite{Stormer99}, to spinons in antiferromagnetic chains \cite{Sachdev11} and, more recently, fractionalisation of Dirac electrons in graphene \cite{Bolotin09} and magnetic monopoles in spin ice \cite{Castelnovo08}.
Even more surprising is the fragmentation of the degrees of freedom themselves, leading to coexisting and a priori independent ground states. This puzzling phenomenon was recently put forward in the context of spin ice, in which the magnetic moment field can fragment, resulting in a dual ground state consisting of a fluctuating spin liquid, a so-called Coulomb phase \cite{Henley10}, on top of a magnetic monopole crystal \cite{Bartlett14}.
Here we show, by means of neutron scattering measurements, that such fragmentation occurs in the spin ice candidate \ndzr. We observe the spectacular coexistence of an antiferromagnetic order induced by the monopole crystallisation and a fluctuating state with ferromagnetic correlations. Experimentally, this fragmentation manifests itself via the superposition of magnetic Bragg peaks, characteristic of the ordered phase, and a pinch point pattern, characteristic of the Coulomb phase.
These results highlight the relevance of the fragmentation concept to describe the physics of systems that are simultaneously ordered and fluctuating.}

\begin{figure}[!]
\includegraphics[height=13cm]{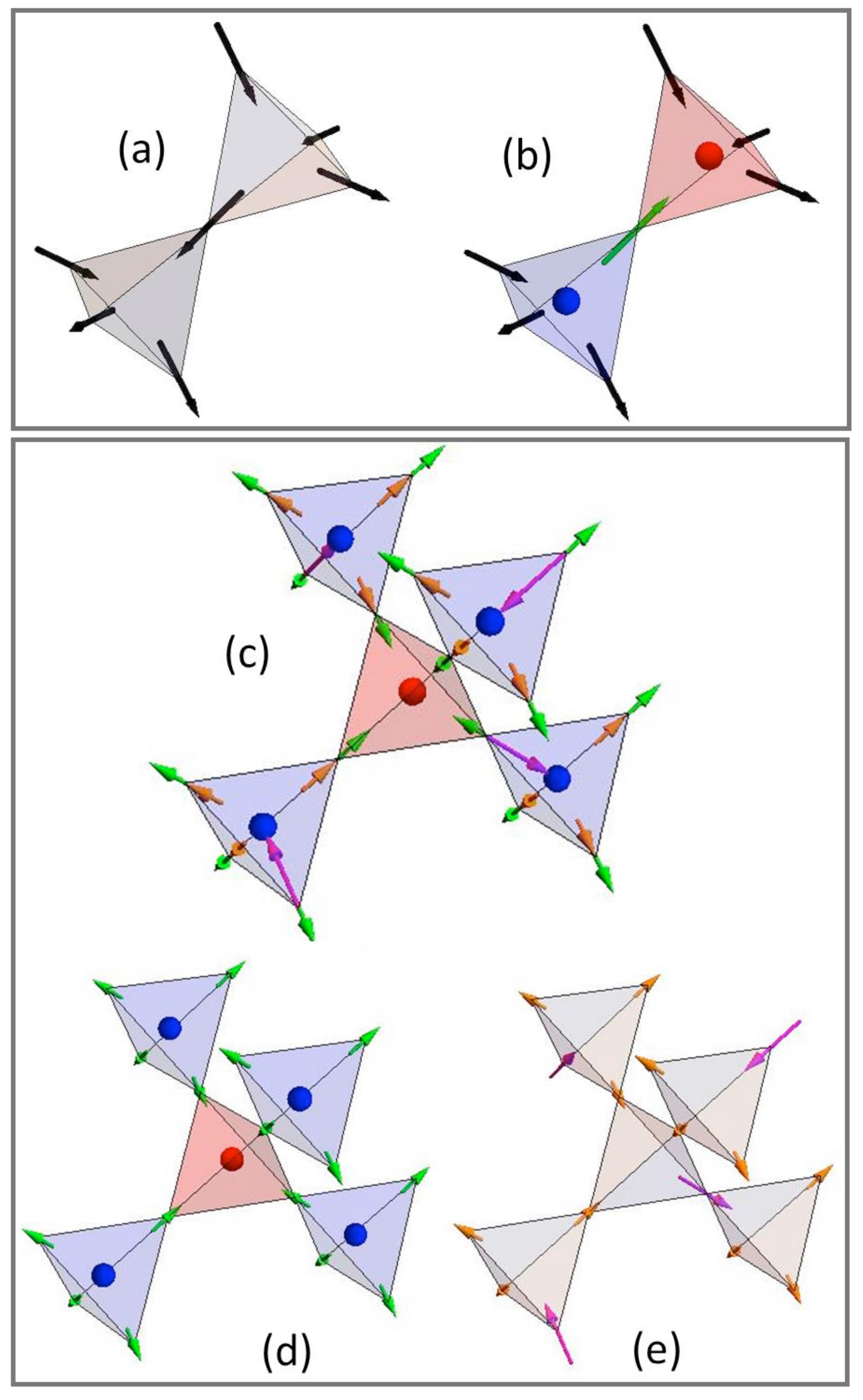}
\setstretch{1}
\caption{\label{fig1} {\bf Spin ice and magnetic fragmentation.} Top panel: (a) Schematic of tetrahedra obeying the ice rule. The spin state $\left\{s_1,s_2,s_3,s_4\right\}$ of a tetrahedron can be described using the convention: $s_i=1$ for a spin pointing in and $-1$ for a spin pointing out. The ice rule is simply written as $\sum_i s_i = 0$, which corresponds to a divergence free condition. 
(b) A spin-flip generates two magnetic monopoles. 
The red (blue) monopole in (b) can be written $m_+=\left\{1,1,1,-1\right\}$ ($m_-=\left\{-1,-1,-1,1\right\}$), so that $\sum_i s_i = \pm 2$ on a tetrahedron.  
Bottom panel: in the fragmentation theory, the local magnetic moment field fragments through a Helmholtz decomposition into two parts, a divergence full and a divergence free part. Each fragment carries components $s_i \neq \pm 1$. 
For instance, for the $m_+$ monopole $\left\{1,1,1,-1\right\}$, this leads to: $\left\{1,1,1,-1\right\}= \left\{\frac{1}{2},\frac{1}{2},\frac{1}{2},\frac{1}{2} \right\}+\left\{\frac{1}{2},\frac{1}{2},\frac{1}{2},-\frac{3}{2}\right\}$. 
(c) Sketch of monopole crystallisation, with the representation of the fragmented moments. 
(d) The divergence full contributions (green arrows) $\left\{\frac{1}{2},\frac{1}{2},\frac{1}{2},\frac{1}{2}\right\}$ and $\left\{-\frac{1}{2},-\frac{1}{2},-\frac{1}{2},-\frac{1}{2}\right\}$, form an ``all in - all out" state and carry the magnetic charge ($\sum_i s_i = \pm 2$). 
(e) The second contribution is made of three components $1/2$ (orange arrows) and one component $3/2$ (magenta) (for example $\left\{\frac{1}{2},\frac{1}{2},\frac{1}{2},-\frac{3}{2}\right\}$), in such a way that the $\sum_i s_i = 0$ constraint is fulfilled. It thus obeys a divergence free condition, but is different from the ice rule, 
and forms a new kind of Coulomb phase.
The same argument can be made for the other magnetic configurations $m_+=\left\{1,1,-1,1\right\}, \left\{1,-1,1,1\right\}$, or $\left\{-1,1,1,1\right\}$ (and associated $m_-$). The divergence full contribution remains identical while the position of the magenta $3/2$ component that emerges in the divergence free contribution changes, as a fingerprint of the degeneracy.}
\end{figure}

\vspace{0.3 cm}
The physics of spin ice materials is intimately connected with the pyrochlore lattice, made of corner sharing tetrahedra. On the corners of these tetrahedra reside rare-earth magnetic moments $\vec{J}_i$ of size $\mu$, which, as a consequence of the strong crystal electric field, are constrained to point along their local trigonal axes $\vec{z}_i$, and behave like Ising spins.  
The magnetic interactions are made of nearest-neighbour exchange ${\cal J}$ and dipolar interactions \cite{denHertog00}: 
\begin{equation}
{\cal H} = {\cal J} \sum_{<i,j>}{\vec{J}_i \cdot \vec{J}_j }+ {\cal D} r_{\rm nn}^3 \sum_{<i,j>}{ \left \lbrack \frac{\vec{J}_i \cdot \vec{J}_j}{r_{ij}^3} - \frac{3(\vec{J}_i \cdot \vec{r}_{ij})(\vec{J}_j \cdot \vec{r}_{ij})}{r_{ij}^5}\right \rbrack}
\label{h}
\end{equation}
where $ {\cal D}= \dfrac{\mu_o (g_{\rm J} \mu_{\rm B})^2}{4\pi r_{\rm nn}^3}$ and $r_{\rm nn}$ is the nearest neighbour distance between rare-earth ions. The nearest neighbour spin ice Hamiltonian is obtained by truncating the Hamiltonian (\ref{h}), yielding:
\begin{equation}
{\cal H}_{\rm nn} =  \frac{-{\cal J} + 5 {\cal D}}{3}  ~ \sum_{<i,j>} J^z_i J^z_j.
\label{h2}
\end{equation}
When the effective interaction ${\cal J}_{\rm eff} = \left(-{\cal J} + 5 {\cal D} \right)/3$ is positive, i.e. when the dipolar term overcomes the antiferromagnetic exchange, a very unusual magnetic state develops, the spin ice state.
The system remains in a highly correlated but disordered ground state where the local magnetisation fulfills the so-called ``ice rule": each tetrahedron has two spins pointing in and two out (see Figure \ref{fig1}(a)), in close analogy with the rule which controls the hydrogen position in water ice \cite{Harris97}. The extensive degeneracy of this ground state results in a residual entropy at low temperature which is well-approximated by the Pauling entropy for water ice \cite{Ramirez99}. 

Such highly degenerate states where the organising principle is dictated by a local constraint belong to the class of Coulomb phases \cite{Henley10,Huse03,Bergmann06}: the constraint (the ice rule for spin ice), can be interpreted as a divergence free condition of an emergent gauge field. This field has correlations that fall off with distance like the dipolar interaction \cite{Isakov04, Henley05}. In reciprocal space, this power-law character leads to bow tie singularities, called pinch points, in the magnetic structure factor. They form a key experimental signature of the Coulomb phase physics. They have been observed by neutron diffraction in the spin ice materials \hoti\ and \dyti\,, in excellent agreement with theoretical predictions \cite{Fennell09, Morris09}. 

Classical excitations above the spin ice manifold are defects that locally violate the ice rule and so the divergence free condition: by reversing the orientation of a moment, ``three in - one out" and ``one in - three out" configurations are created (see Figure \ref{fig1}(b)). Considering the Ising spins as dumbbells with two opposite magnetic charges at their extremities, such defects result in a magnetic charge in the centre of the tetrahedron, called a magnetic monopole, that give rise to a non-zero divergence of the local magnetisation \cite{Castelnovo08}.

Recently, theoreticians have introduced the concept of magnetic moment fragmentation \cite{Bartlett14}, whereby the local magnetic moment field fragments into the sum of two parts, a divergence full and a divergence free part (see Figure \ref{fig1}(c)): for example, a monopole in the spin configuration $m=\left\{1,1,1,-1\right\}$ on a tetrahedron can be written $m=\frac{1}{2} \left\{1,1,1,1\right\} + \frac{1}{2} \left\{1,1,1,-3\right\}$. In this decomposition, the first term carries the total magnetic charge of the monopole.
If the monopoles organise as a crystal of alternating magnetic charges, the fragmentation leads to the superposition of an ordered ``all in - all out" configuration (Figure \ref{fig1}(d)) and of an emergent Coulomb phase associated with the divergence free contribution (Figure \ref{fig1}(e)).

\begin{figure}[t!]
\includegraphics[width=8cm]{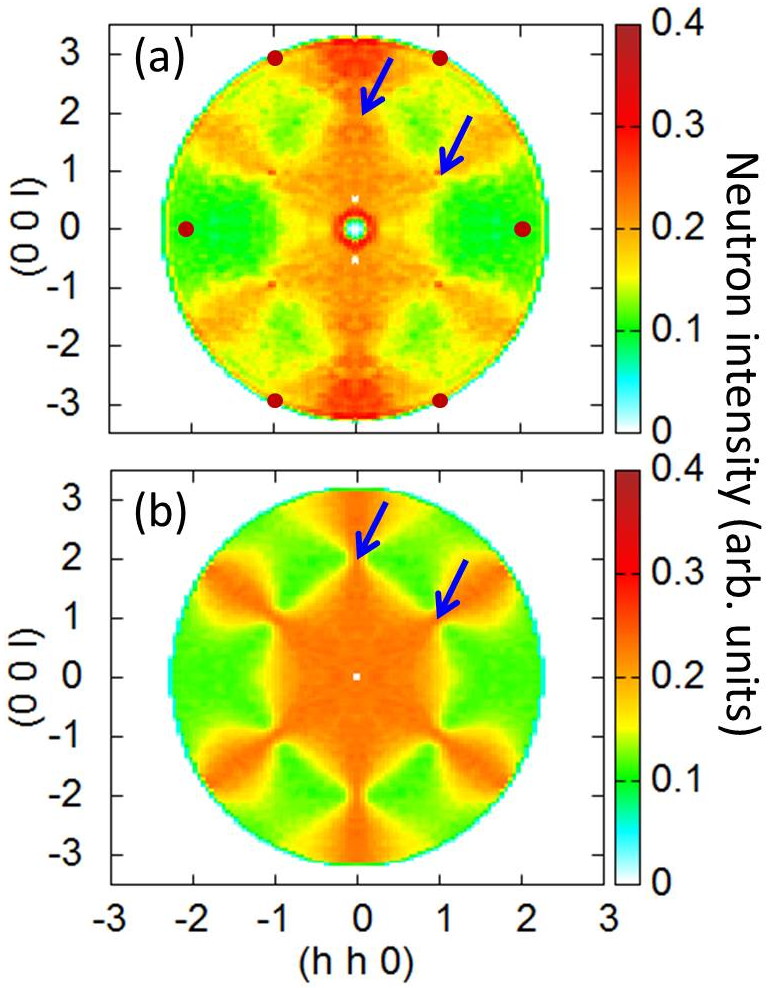}
\caption{\label{fig2} {\bf Pinch point pattern in \ndzr.} (a) Inelastic neutron scattering intensity averaged in the energy range $45 < E < 55~\mu$eV measured at $T=60$ mK with an incident wavelength $\lambda=6$~\AA. The red spots denote the antiferromagnetic Bragg peak positions that appear at zero energy transfer. 
(b) Dynamical structure factor $S(Q,E=50~\mu$eV$)$ calculated in the RPA approximation for the pseudo spin 1/2 model described by equation \ref{h3}, with ${\cal J'}= 1.2$ K and ${\cal K}= -0.55$ K. The model takes into account a ferromagnetic exchange ${\cal J'}$ along with a transverse interaction ${\cal K}$ whose physical origin is a coupling between octopoles. Blue arrows indicate the pinch point positions.}
\end{figure}

\begin{figure*}[t!]
\includegraphics[width=\textwidth]{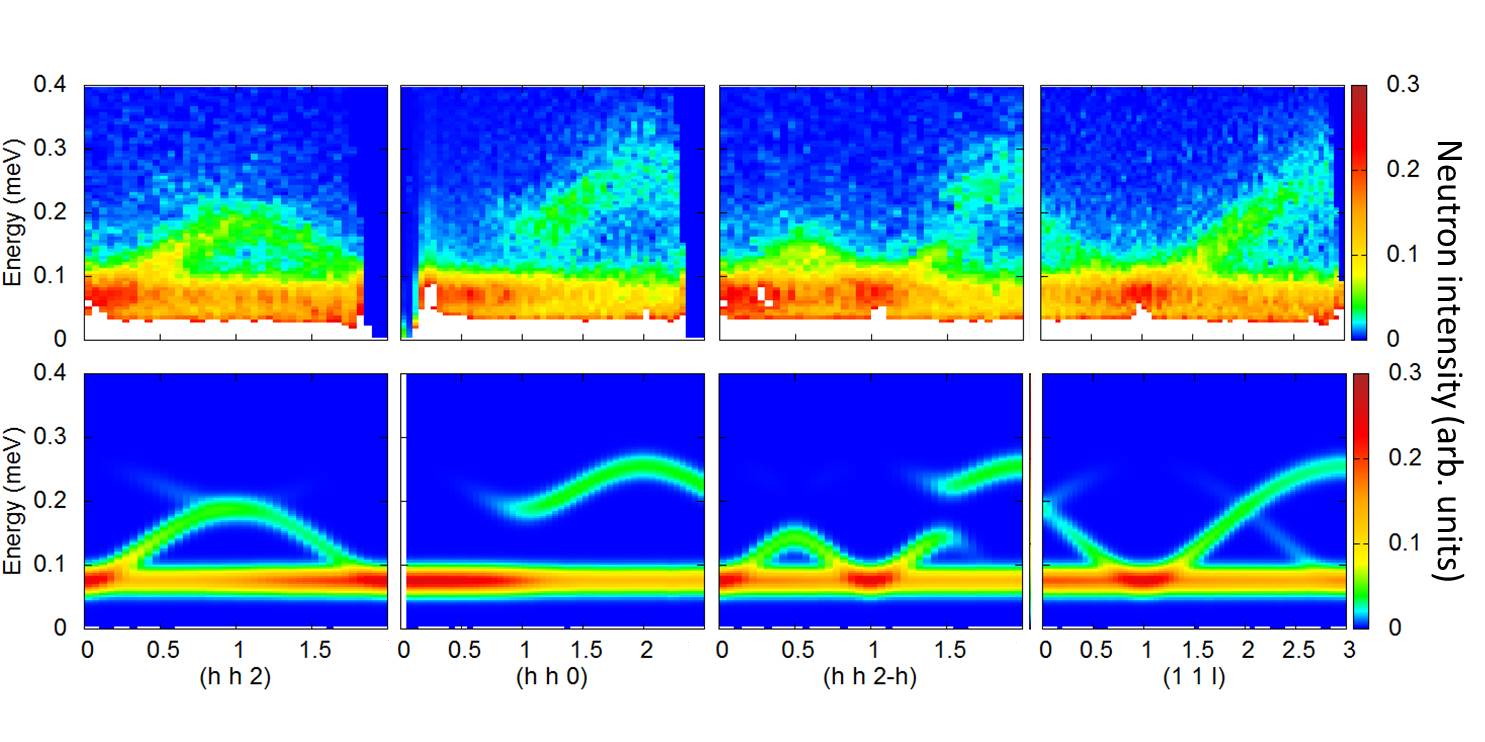}
\caption{\label{fig3} {\bf Magnetic excitation spectrum.} Top: Inelastic neutron scattering spectra taken at 60~mK along various high symmetry directions with an incident wavelength $\lambda=6$ \AA. Bottom: RPA calculation of the corresponding spectra for the pseudo spin $1/2$ model described by equation \ref{h3}, with ${\cal J'}=1.2$ K and ${\cal K}= - 0.55$ K. Note that the same dispersion curve is obtained with a ferro octopolar coupling ${\cal K} = +1.65$ K: the sign of the octopolar coupling does not affect the physics of the spin model. The factor of three between the antiferro and ferro octopolar coefficients is due to the molecular field amplitude of the octopolar phase, the ``all in - all-out'' molecular field being three times larger than the ``two in - two out'' one.}
\end{figure*}

This monopole crystallisation occurs when the monopole density is high enough so that the Coulomb interaction between monopoles (which originates in the dipolar interaction between magnetic moments) is minimised through charge ordering, while the remaining fluctuating divergence free part provides a gain in entropy. 

The necessary conditions for an experimental realisation of this physics are severe:
 in pyrochlore systems, the fragmentation is expected in the case of strong Ising anisotropy combined with effective ferromagnetic interactions, and for a specific ratio between the dipolar and exchange interactions in order to form the  crystal of monopoles. 
If fragmentation occurs, the theory predicts that the magnetic structure factor should exhibit both Bragg peaks characteristic of the ``all in - all out" structure and a pinch point pattern typical of a Coulomb phase \cite{Bartlett14}.

The pyrochlore system \ndzr\, is a good candidate in the search for such a system. Previous studies have provided evidence for the strong Ising character of the \nd\, ion, and for ferromagnetic interactions, inferred from the positive Curie-Weiss temperature $\theta_{\mathrm{CW}}=195$~mK \cite{Ciomaga15}. Moreover, \ndzr\ orders below $T_{\mathrm N}=285$~mK in an ``all in - all out" state carrying a reduced ordered magnetic moment of about one third of the total Nd$^{3+}$ magnetic moment $\mu_{\rm eff}=2.4~\mu_{\mathrm B}$ \cite{Lhotel15} (See Supplementary Information). 

To demonstrate that the fragmentation occurs in \ndzr, it is essential to observe signatures of the Coulomb phase. To this end, neutron scattering experiments have been carried out as a function of temperature and magnetic field on a large single crystal. 
As shown in Figure \ref{fig2}(a), the key point here is that the neutron data do exhibit at 60 mK arm like features along the $(00\ell)$ and $(hhh)$ directions, with pinch points at the $(002)$ and $(111)$ positions, expected in the Coulomb phase \cite{Bartlett14}.
This pinch point pattern is observed simultaneously with the ``all in - all out" Bragg peaks \cite{Ferey86, Lhotel15} at $(220)$ and $(113)$, which we interpret as evidence for fragmentation and monopole crystallisation.

\begin{figure}[!]
\includegraphics[width=8cm]{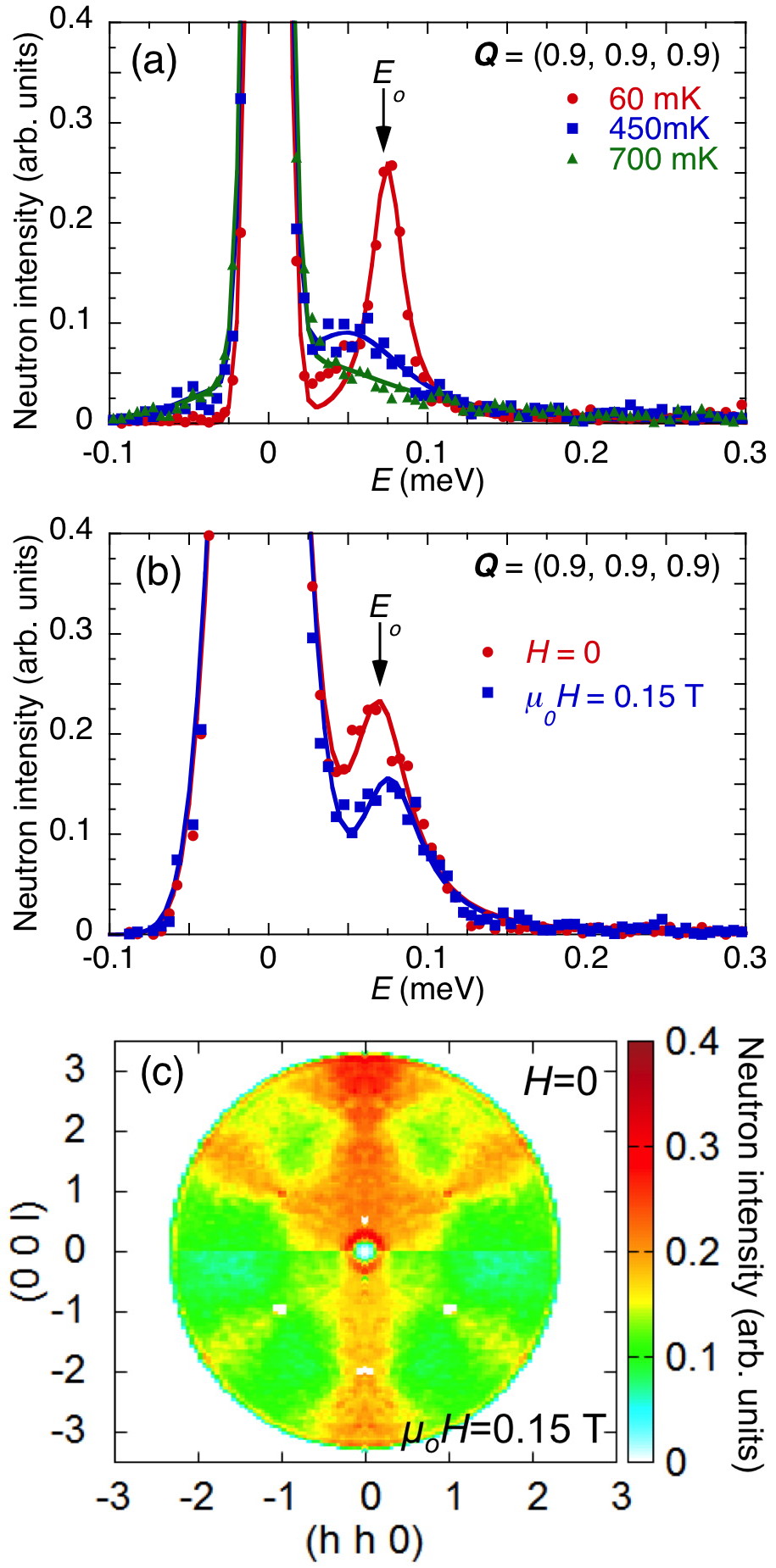}
\caption{\label{fig4} {\bf Temperature and field dependence of the pinch point pattern.} Excitation spectra at ${\bf Q}=(0.9, 0.9,0.9)$ (a) at several temperatures for $\lambda=8.5$ \AA, (b) in zero field and in a 0.15 T applied field in the [1 -1 0] direction at 60 mK, for $\lambda=6$ \AA. Above the strong elastic incoherent scattering a flat mode appears at $E_o$ whose structure factor exhibits the pinch point pattern. Note that due to the different energy resolution at $\lambda=6$ \AA, the peak appears broader in (b).
(c) Inelastic neutron scattering map measured at $T=60$ mK and averaged in the energy range $45 < E < 55~\mu$eV showing a similar pattern but different intensity between zero field (top) and 0.15 T (bottom) data.}
\end{figure}

Importantly, this structured neutron scattering signal appears as a flat mode at finite energy around $E_o=70~\mu$eV (see Figures \ref{fig3} top and \ref{fig4}(a) and (b)). 
In addition, above this flat mode, collective dispersive excitations stem from the pinch points and not from the antiferromagnetic ``all in - all out" wave vectors (see Figure \ref{fig3} top). They are characterised by a spin gap  $\Delta \approx E_o$ and reach a maximum energy of about 0.25 meV. 

When increasing the temperature, the pinch point pattern and the collective modes persist up to 600~mK, far above the antiferromagnetic ordering ($T_{\mathrm N}=285$~mK) (see Supplementary Information). While the energy gap, and the intensity of these features, decrease as the temperature increases (see Figure \ref{fig4}(a)), the energy range of the dispersion remains unaffected up to 450~mK. 
This temperature dependence suggests a scenario in which the fragmentation takes place well above $T_{\rm N}$: at the temperature where the ferromagnetic correlations start to develop, a Coulomb phase arises in coexistence with a liquid of monopoles. The latter finally crystallises upon cooling in an ``all in - all out" phase at $T_{\rm N}$, leaving the Coulomb phase unchanged. 
The field dependence is consistent with this scenario  (see Figure \ref{fig4}(b) and (c)): at an applied field of 0.15 T, where magnetisation measurements show that the ``all in - all out" state is replaced by a field induced ordered state \cite{Lhotel15}, the Coulomb phase characteristics remain, albeit with less intensity. This observation further confirms the fragmentation scenario in which the divergence free and full parts of the magnetic moment field behave independently.  

This peculiar spin dynamics, and especially the existence of dispersive modes, are puzzling in an Ising-like system. They call for the existence of additional transverse terms in the Hamiltonian given in equation \ref{h2}. To address this point, the magnetic moments should not be considered as Ising variables, but as pseudo spin half $\sigma_i$ spanning the $|\uparrow \downarrow \rangle$ crystalline electric field (CEF) doublet states. 

Considering the very peculiar ``dipolar-octopolar" nature of the Kramers \nd\, doublet \cite{Huang14,Lhotel15,Abragam70,Watahiki11} such transverse terms arise from a coupling between octopolar moments. Indeed, while $\langle \uparrow | \vec{J} | \downarrow \rangle \equiv 0$ because of those CEF properties, it can be shown using the explicit wave functions determined in Ref. \citenum{Lhotel15} that the octopole ${\cal T} = i(J^+ J^+J^+ - J^- J^-J^-)$ is the relevant operator since $\langle \uparrow | {\cal T} | \downarrow \rangle \ne 0$. Introducing an octopole-octopole coupling ${\cal V} = \sum_{<i,j>} \kappa~{\cal T}_i~{\cal T}_j$, where $\kappa$ denotes the strength of the octopolar coupling, and projecting it onto the pseudo spin $1/2$ subspace leads to:

\begin{equation}
{\cal H}_{1/2} =\sum_{<i,j>} {\cal J'}\sigma^z_i \sigma^z_j  + {\cal K} \sigma^y_i \sigma^y_j 
\label{h3}
\end{equation}
where $\sigma^{y,z}$ are the pseudo spin components in the local coordinates, ${\cal J'}$ is an effective exchange interaction ${\cal J'}=(g_z/g_{\rm J})^2 {\cal J}_{\rm eff}$ and ${\cal K} = 4 \kappa |\langle \uparrow | {\cal T} | \downarrow \rangle|^2$. For \ndzr, $g_{\rm J}=8/11$ and $g_x=g_y=0, g_z=4.5$. 

The Hamiltonian parameters ${\cal J'}$ and ${\cal K}$ can be estimated by fitting the inelastic neutron scattering spectra. From calculations in the Random Phase Approximation (RPA) \cite {jensen,kao03,petit14,Robert15} (see Supplementary Information), it is found that the bandwidth of the collective modes is related to ${\cal J'}$ while the shift of the pinch point pattern up to $E_o$ is induced by the transverse term ${\cal K}$. This is reminiscent of the role of the antisymmetric Dzyaloshinskii-Moriya interaction which lifts the ``weathervane" flat mode in kagome systems up to finite energy \cite{Matan06}. Such transverse terms might also be at the origin of the inelastic pattern observed in the quantum spin ice candidate \przr \cite{Kimura13}. The best agreement is obtained for ${\cal J'}=1.2$~K and ${\cal K}=-0.55$ K (see Figures \ref{fig2}(b) and \ref{fig3} bottom). For these values, the RPA ground state is an ordered octopolar phase. It is worth noting that, while this RPA calculation accounts for the behaviour of the divergence free part of the magnetic moment, it is unable to capture the fragmentation mechanism.

We have thus shown that the predicted fragmentation process \cite{Bartlett14} exists in the spin ice material \ndzr. Below 700 mK, the magnetic moment field fragments into two parts: a divergence full part which crystallises at $T_{\rm N}=285$ mK, and a divergence free part for which transverse terms induce gapped and dispersive excitations. Our results highlight that the two fragments behave independently as a function of field and temperature, which opens the appealing possibility of manipulating them separately. 

Beyond the classical fragmentation theory described in Ref. \citenum{Bartlett14}, the importance of transverse terms to describe our observations emphasises the need for considering quantum effects in further theoretical studies. Indeed, in the classical scheme, the crystallisation occurs when the energy required to create the assembly of fragmented monopoles is balanced by the repulsive energy among them and thus depends on the competition between exchange and dipolar interactions. In the present case, transverse octopolar couplings might enhance the interactions between monopoles thus promoting their crystallisation. We thus anticipate that our experiment will pave the way towards a quantum theory of fragmentation, involving such tranverse terms.  

In a broader context, the fragmentation theory relies on the Helmholtz decomposition of a charged field, widely used to describe continuous fluid media in a large variety of fields, from fluid mechanics to robotics \cite{Bhatia13}.
This decomposition allows to identify new relevant degrees of freedom, which could not have been separated otherwise. Our results in \ndzr\ point out its applicability to describe, more generally, localized moment systems where fluctuating and ordered phases coexist. This might cover the case of the pyrochlore compound \ybti, a system showing a strongly reduced ferromagnetic ordering \cite{chang} and a peculiar fluctuation spectrum, and whose physics is likely governed by competing phases \cite{Robert15, Jaubert15}.
In spin ice, the Helmholtz decomposition is applied at a microscopic level on the emergent gauge field of the Coulomb phase and on its charges, the monopoles. Our experimental findings give a concrete form to these concepts. The observation of fragmentation in \ndzr\ will thus stimulate new conceptual approaches in physical systems where such emergent fields exist.

\section*{Methods}

Single crystals of \ndzr\, were grown by the floating-zone technique using a four-mirror xenon arc lamp optical image furnace \cite{monica1,Ciomaga15}. \\
Inelastic neutron scattering experiments were carried out at the Institute Laue Langevin (ILL, France) on the IN5 disk chopper time of flight spectrometer and operated with $\lambda~=~8.5$~\AA\ or $\lambda=6$ \AA. The \ndzr\, single crystal sample was attached to the cold finger of a dilution insert and the field was applied along [1 -1 0]. 
The data were processed with the {\it Horace} software, transforming the recorded time of flight, sample rotation and scattering angle into energy transfer and $Q$-wave-vectors. \\
The neutron diffraction data were taken at the D23 single crystal diffractometer (CEA-CRG, ILL France) with a copper monochromator and using $\lambda=1.28$ \AA. Here the field was applied along the $[111]$ direction. Refinements were carried out with the Fullprof software suite \cite{fullprof}.\\
The magnetic diffuse scattering was measured on the D7 diffractometer installed at the ILL, with $\lambda=4.85$ \AA, using standard polarization analysis technique with the guiding field along the vertical axis [1 -1 0]. \\
Calculations are carried out on the basis of a mean field treatment of a Hamiltonian taking into account the dipolar exchange as well as an octopolar coupling between the CEF ground doublet states of the \nd\ ion. This Hamiltonian is written in terms of a pseudo spin $1/2$ spanning these states. The spin dynamics is then calculated numerically in the Random Phase Approximation \cite{jensen,kao03,petit14,Robert15}.\\
More details are provided in the Supplementary Information.

\section*{Acknowledgements}
We acknowledge the ILL for the beam time. We also thank P.C.W. Holdsworth, V. Simonet and Y. Sidis for fruitful discussions. MCH, MRL, and GB acknowledge financial support from the EPSRC, United Kingdom, Grant No. EP/M028771/1.\\



%

\newpage

\renewcommand{\figurename}{FIG. S}
\renewcommand{\tablename}{TABLE S}
\renewcommand{\theequation}{\alph{equation}}

\makeatletter
\renewcommand{\@biblabel}[1]{\quad S#1. }
\makeatother

 \setcounter{figure}{0} 

\onecolumngrid
\begin{center} {\bf \large Observation of magnetic fragmentation in spin ice \\
 Supplementary Information} \end{center}
\vspace{0.5cm}
\twocolumngrid

\maketitle

\section{Methods} 
\subsection{Sample preparation}

Single crystals of \ndzr\, were grown by the floating-zone technique using a four-mirror xenon arc lamp optical image furnace (CSI FZ-T-12000-X VI- VP, Crystal Systems, Inc., Japan) $^{\rm S}$\cite{monica1b}$^{,\rm S}$\cite{Ciomaga15b}. The growths were carried out in air at ambient pressure and at growth speeds in the range 10 - 15 mm/h. The two rods (feed and seed) were counter-rotated at a rate of 20 - 30 rpm. The crystals were aligned using a Laue X-ray imaging system with a Photonic-Science Laue camera.

\subsection{Neutron scattering experiments}

Inelastic neutron scattering experiments were carried out on the IN5 disk chopper time of flight spectrometer operated by the Institute Laue Langevin (France). Different wavelengths $\lambda=8.5$ \AA\ and $\lambda=6$ \AA\ were used depending on the desired energy resolution (about 20 and 40 $\mu$eV respectively). The \ndzr\, single crystal sample was attached to the cold finger of a dilution insert, and placed in a vertical magnet with a maximum field of 2.5 T. The sample was mounted in order to have the $(hh0)$ and $(00\ell)$ reciprocal directions in the horizontal scattering plane. The field was thus applied along [1 -1~0]. The data were processed with the {\it Horace} software$^{\rm S}$\cite{horace}, transforming the recorded time of flight, sample rotation and scattering angle into energy transfer and $Q$-wave-vectors. The offset of the sample rotation was determined based on the Bragg peaks positions. In all experiments, the sample was rotated by steps of 1 degree, covering a total range of 130 degrees. The counting time was about 10 minutes per angle value. When appropriate, the data have been symmetrised.

The neutron diffraction data were collected using the D23 single crystal diffractometer (CEA-CRG, ILL France) operated with a copper monochromator and using $\lambda=1.28$ \AA. The sample was mounted on a dilution stick placed in a 6 T vertical cryomagnet, this vertical direction corresponding to the $[111]$ direction of the crystal. Refinements were carried out with the Fullprof software suite$^{\rm S}$\cite{fullprofb}.

The magnetic diffuse scattering was measured on the D7 diffractometer installed at ILL, with a wavelength $\lambda=4.85$~\AA\ and using standard polarization analysis technique with the guiding field along the vertical axis [1 -1 0] of the sample. Maps at 10 K have been subtracted from the low temperature data to eliminate the background. Those maps are consistent with the IN5 data. Note that the diffraction technique measures instantaneous correlations, integrating the dynamical response observed at IN5 up to an energy transfer equal to 3.5 meV.

\section{Magnetic structure}

\begin{figure}[h!]
\includegraphics[width=9cm]{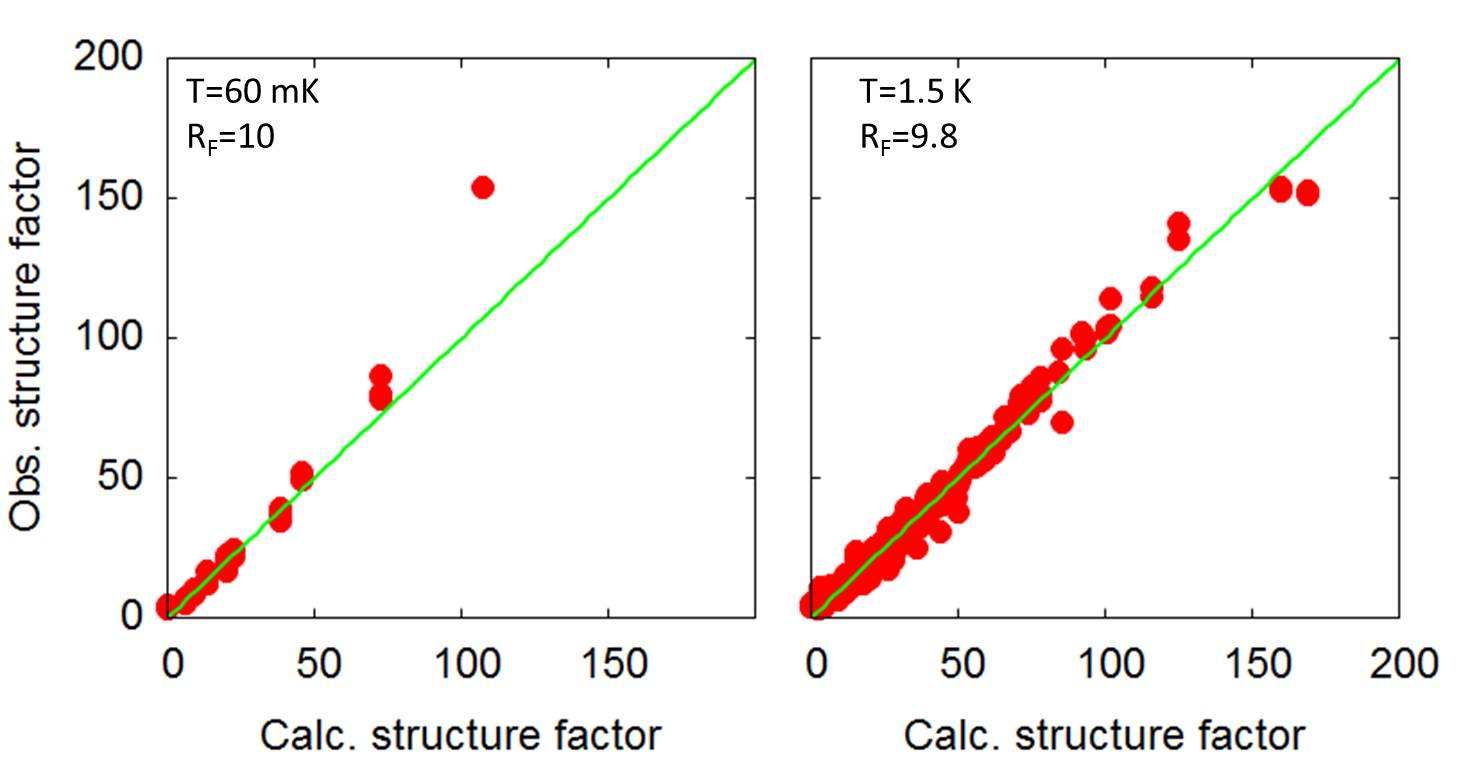}
\caption{\label{fig1-supmat} (Left) Joint refinement of the crystal and magnetic neutron structure factors at 60 mK. (Right) Crystal refinement at 1.5 K.}
\end{figure}

A series of Bragg peak intensities was collected at 1.5~K and 50 mK and for various fields applied up to 8 T. At 1.5~K, the data are in agreement with the published pyrochlore structure ({\it Fd$\bar{3}$m} space group) with the 48f oxygen positions determined by the parameter $x=0.337\pm0.002$~ $^{\rm S}$\cite{Ciomaga15b}. The magnetic contribution increases below $T_{\rm N}$ on top of those crystalline peaks. The Fullprof refinement of the magnetic structure confirms prior powder diffraction results$^{\rm S}$\cite{Lhotel15b} and shows that \ndzr\, orders in the ``all in - all-out'' (AIAO) structure, see Figure S\ref{fig1-supmat}. The refined ordered \nd\, magnetic moment is ($0.8\pm0.1) \mu_{\rm B}$.

It is worth noting that the transition temperature of the N\'eel order depends on the nature of the sample: it is slightly larger ($T_{\rm N} \sim 400$ mK) in the powder samples$^{\rm S}$\cite{Lhotel15b}$^{,\rm S}$\cite{Xu15} than in the single crystal ($T_{\rm N} \sim 285$ mK). A similar behaviour has been observed in other pyrochlore systems such 
as \ybti\ $^{\rm S}$\cite{ybti}, and was ascribed to the crystalline quality of the samples, which is supposed to be better in polycrystalline samples. In the studies we have performed, despite a different N\'eel temperature, the magnetic properties of the powder sample and single crystal are similar, especially the existence of a partial ordered moment. The presence of structural defects might affect the fragmentation process and thus the ordering temperature. Nevertheless, the scattering pattern characteristic of ice-like correlations persists up to 600 mK, 
far above the N\'eel temperature of all samples. 

\section{Temperature dependence}
The maps performed on D7 allow to observe both the spin-ice pattern along with the AIAO Bragg peaks, especially (220) and (113).

\begin{figure}[h!]
\includegraphics[width=9cm]{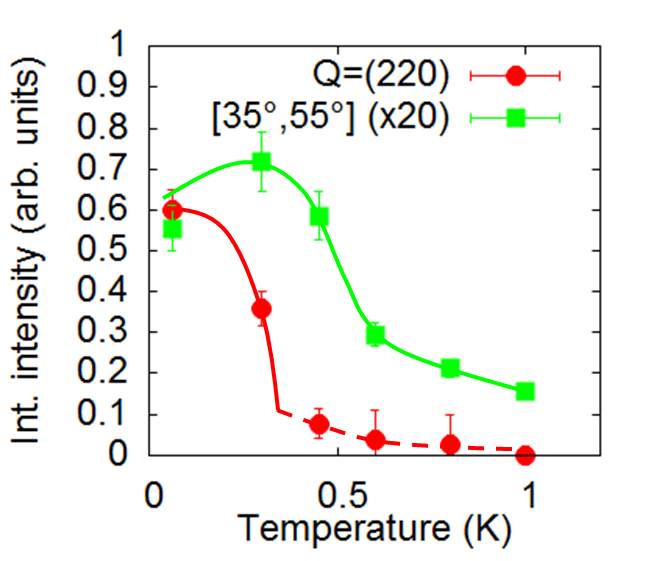}
\caption{\label{figD7-supmat}Red dots: Averaged neutron intensity around the (220) Bragg peak as a function of temperature. Green squares: Averaged neutron intensity in a sector delineating the arm along $(hhh)$ of the spin-ice pattern. This sector is defined in the range 35 to 55 degrees.}
\end{figure}

Figure \ref{figD7-supmat} shows the magnetic neutron intensity integrated over a small region around (220) as well as over a sector delineating the arm along $(hhh)$ (i.e. the direction of the (111) pinch point), between 35 and 55 degrees. These data show that the Bragg intensity disappears at $T_{\rm N}$, between 300 and 450 mK, while the spin-ice pattern persists above $T_{\rm N}$, up to 600 mK and becomes barely observable only above 750 mK.

Interestingly, the intensity of the arms seems to be maximum around 300 mK. Further analysis is in progress to confirm this effect, and understand it. 

\begin{figure*}[t]
\includegraphics[width=\textwidth]{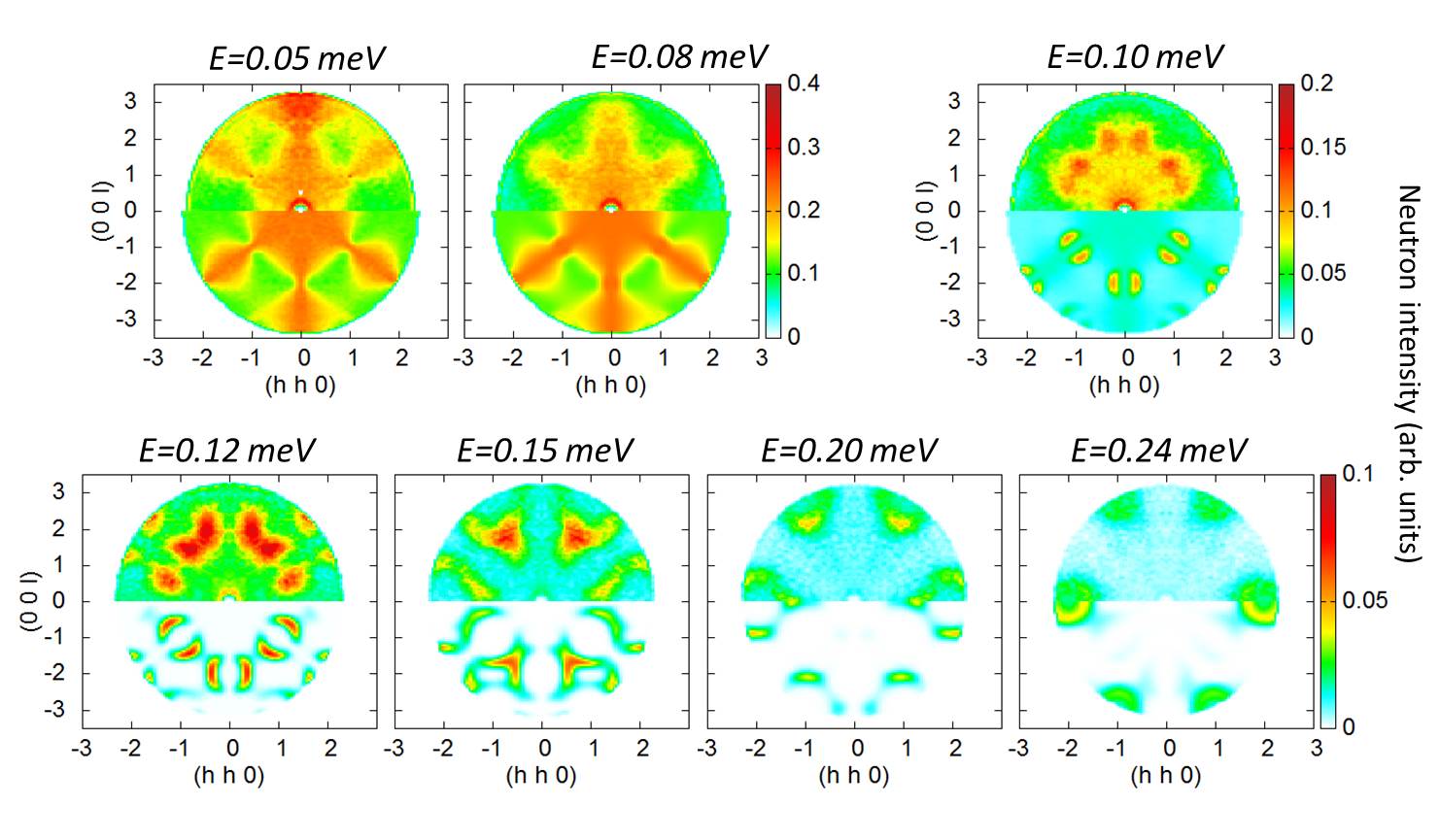}
\caption{\label{fig2-supmat}(Top) Inelastic neutron scattering spectra taken at 60 mK at different constant energies ranging from 0.06 to 0.24~meV, with an incident wavelength $\lambda=6$ \AA. (bottom) Corresponding RPA calculations for the pseudo spin $1/2$ model given by Eq. (\ref{h12}) with ${\cal J'}=1.2$~K and ${\cal K}=-0.55$~K. The calculations have been normalized to the data by the same factor above 0.1~meV. Below this value, a larger normalization factor was used to allow a better comparison. Indeed, the intensity of the cuts taken at small energies, is articially enhanced because of the energy resolution which integrates part of the inelastic line.}
\end{figure*}

\section{Inelastic neutron scattering data}

The main text of the present paper shows maps as a function of energy transfer and $Q$ along particular directions of reciprocal space. For the sake of completeness, Figure S\ref{fig2-supmat} shows constant energy cuts covering the full bandwidth of the spectrum. Above the inelastic line, a flat mode appears at around $E_o=70~\mu$eV, with a structure factor characterised by the pinch point pattern. Above this mode, collective excitations disperse from the $(111)$ and $(002)$ $Q$-positions. 

The spin ice pattern and the pinch points are more clearly observed for energies slightly below $E_o$; indeed, at $E_o$, they are blurred by the contribution that arises from the dispersing branches because of the energy resolution.
With increasing energy, the top of the band appears at the $(113)$ and $(220)$-like $Q$-positions. Figure S\ref{fig2-supmat} also shows the corresponding calculations carried out in the model given by Eq. (\ref{h12}) (see below).

\section{Fields and Helmholtz decomposition}

The decomposition of a vector field into divergence free and divergence full parts is called the Helmholtz decomposition. It is very general and is commonly used in fluid mechanics and electromagnetism$^{\rm S}$\cite{Bhatia13b}. More precisely, it states that a vector field $\vec{V}$ can be written $\vec{V} = \vec{\nabla} \times \vec{A} - \vec{\nabla} \psi$ where $ \vec{A}$ and $\psi$ are vector and scalar fields respectively. As $\vec{\nabla} . (\vec{\nabla} \times \vec{A}) \equiv 0$, the first term of this decomposition is divergence free, while the second is not, with $\vec{\nabla} . \vec{\nabla} \psi = \Delta \psi$ but is curl free $\vec{\nabla} \times \ \vec{\nabla} \psi \equiv 0$.\\

In the context of Ising spins on the pyrochlore lattice, the spin state of a tetrahedron can be described by $\left\{s_1,s_2,s_3,s_4\right\}$ using the convention $s_i=1$ for a spin pointing in and $-1$ for a spin pointing out. The divergence free condition can be expressed as $\sum_{i=1,2,3,4} s_i= 0$. \\
As a result, the ``2 in - 2 out" configurations of the spin ice state do not carry any charge and are divergence free. A single magnetic excitation creating a magnetic monopole ``3 in - 1 out"  or ``3 in - 1 out" is charged with $\sum_{i=1,2,3,4} s_i=\pm 2$, while an ``all in - all-out'' tetrahedron is charged with $\sum_{i=1,2,3,4} s_i=\pm 4$. \\

In the fragmentation theory$^{\rm S}$\cite{Bartlett14b}, the vector field of the monopole state is decomposed between an ``all in - all-out'' fragmented field of half amplitude, giving rise to $\sum_{i=1,2,3,4} s_i=\pm 2$ and a divergence free fragment where the sum of the fragmented components is indeed zero (for instance $\frac{1}{2} \left\{1,1,1,-3\right\}$).\\

The Helmholtz decomposition allows one to ``reword" the problem in terms of new fields or new variables. Depending on the Hamiltonian, different regimes can then be expected, where the fragments would behave independently or not. In the classical fragmentation theory described in Ref. S\!\!\citenum{Bartlett14b}, although the crystallisation depends on the competition between exchange and dipolar interactions, it is worth noting that the two fragments are independent.

The present results on \ndzr\, are consistent with this picture: when the first fragment crystallises (the AIAO ordering forms), the second fragment keeps fluctuating and is not markedly affected. We also note that a new and original dynamical response forms: for instance, antiferromagnetic spin waves that could be naively expected to emerge from the AIAO Bragg peaks are not observed. 

The strength of the fragmentation scenario is thus to reveal underlying degrees of freedom, which are not directly visible in the microscopic Hamiltonian. We can thus expect that this theory will be relevant to interpret experimental phenomena, which could not have been explained otherwise, like here in \ndzr, much beyond the field of frustrated magnetism.

\section{Model}
The model developed in the main text is based on the following Hamiltonian for rare earth moments $\vec{J}_i$ at site $i$ of the pyrochlore lattice:
\begin{equation}
{\cal H} = {\cal H}_{\mbox{CEF}} + \frac{1}{2}\sum_{i,j} \vec{J}_i \cdot {\cal J}_{i,j} \cdot \vec{J}_j + \frac{1}{2}\sum_{i,j} {\cal O}_i \cdot \kappa_{i,j} \cdot {\cal O}_j.
\label{hh}
\end{equation}
In this expression, the crystal field (CEF) Hamiltonian ${\cal H}_{\mbox{CEF}}= \sum_{m,n} B_{nm} O_{nm}$ where the $O_{nm}$ are the Wybourne operators ($J=9/2$, $g_J=8/11$ for the \nd\ ion) $^{\rm S}$\cite{wybourne}. ${\cal J}_{i,j}$ is the exchange tensor, ${\cal O}_i$ is a generic term to denote the quadrupolar and octopolar operators, and $\kappa_{i,j}$ is the coupling constant.

The $B_{nm}$ coefficients have been determined in Ref. S\!\!~\citenum{Lhotel15b} and are reproduced in Table S\ref{DO}. The coefficients of the ground doublet wave functions $|\uparrow,\downarrow \rangle$ are typical of a dipolar-octopolar doublet $^{\rm S}$\cite{Abragam70b}$^{,\rm S}$\cite{Huang14b} with the general form :
\begin{eqnarray*}
|\uparrow     \rangle &=& \left(a,0,0,b,0,0,c,0,0,d \right), \\
|\downarrow \rangle &=& \left(d,0,0,-c,0,0,b,0,0,-a \right)
\end{eqnarray*}
associated with the normalisation condition $a^2+b^2+c^2+d^2=1$. Using this explicit formulation, it is possible to show that the projection onto the subspace spanned by $|\uparrow,\downarrow \rangle$ of most dipolar, quadrupolar and octopolar operators is zero, with in particular: 
\begin{eqnarray*}
 J_x & = &  0, \\
 J_y & = &  0, \\
J_x J_y + J_y J_x & = & 0, \\
J_x J_z + J_z J_x & = & 0, \\
J_y J_z + J_z J_y & = & 0.
\end {eqnarray*}
The remaining relevant operators are then
\begin{eqnarray*}
 J_z & = &  \left(
\begin{array}{cc} 
-\mu & v_x \\
v_x & \mu
\end{array}
\right), \\
 J_z^2 & = &  
\left(
\begin{array}{cc} 
\lambda & 0 \\
0 & \lambda
\end{array}
\right), \\
{\cal T} = i (J_+^3-J_-^3) & = &  
\left(
\begin{array}{cc} 
0 & i v_y \\
-i v_y & 0 \\
\end{array}
\right)
\end{eqnarray*}
with
\begin{eqnarray*}
\mu &= & 3/2 (3 a^2 + b^2 - c^2 - 3 d^2) \approx 3.13, \\
v_x & = & 3 b c - 9 a d  \approx  0,\\
\lambda & = & 9/4 (9 (a^2+d^2) + b^2 + c^2) \approx 16.1, \\
v_y & = &  24 (-5 (b^2+c^2) + \sqrt{21} (ac+bd) ) \approx 73.5. 
\end {eqnarray*}
This analysis yields a picture of perfect Ising moments with $g_{\perp} \equiv 0$, $g_ {z}=2g_J\mu = 4.5$. Fluctuations between the two elements of the doublet are, however, induced by $v_x$ and $v_y$. We note that in the present case $v_y$ is very large while $v_x$ is zero. 

\begin{figure*}[!]
\includegraphics[height=7cm]{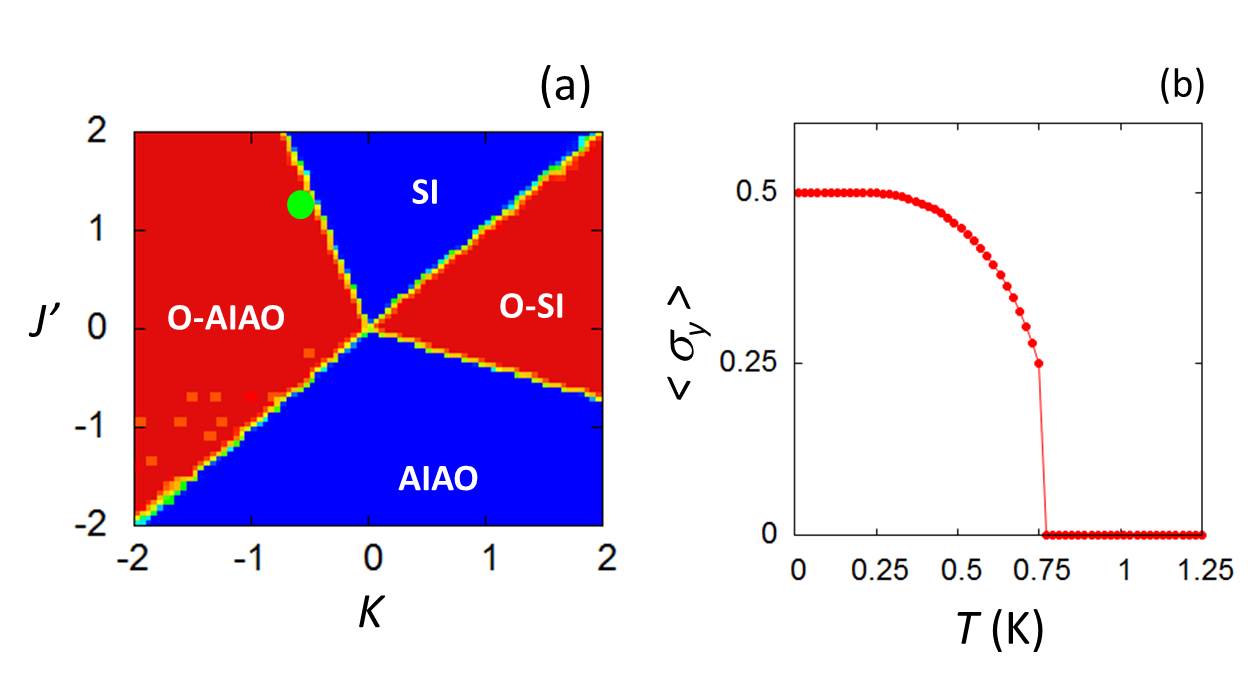}
\caption{\label{fig3-supmat} (a) Phase diagram of the mean field model given by Eq. (\ref{h12}). The AIAO phase is the  ``all in - all out'' phase, ``SI" is an ordered spin ice phase (``2-in 2-out'' phase). The ``O" prefix denotes octopolar phases, either with an ``all in - all out'' nature, the octopolar moments on one tetrahedron being of the same sign, or of the ``SI" nature, 2 octopolar moments being positive, the two others being negative. The magnetic moment is zero in the octopolar phases. The green dot shows the position of the parameters for \ndzr\, determined by the comparison between the experimental and calculated spin dynamics. (b)~Temperature dependence of the octopolar moment ($\sigma_y$ component) in the spin 1/2 mean field model for ${\cal J'}=1.2$~K and ${\cal K}=-0.55$~K. }
\end{figure*}

At this point, it is possible either to deal with the Hamiltonian given by Eq. (\ref{hh}) or to consider a simplified model projected onto the ground state doublet, provided the latter is well protected from the excited CEF states. This issue was solved in Ref. S\!\!\citenum{Huang14b} taking into account symmetry considerations. The most general bilinear Hamiltonian written in terms of the local components of a pseudo spin 1/2 degree of freedom $\sigma$ is of the ``XYZ"  form :
\begin{equation*}
{\cal H}_{1/2} = \frac{1}{2} \sum_{i,j} J^{xx} \sigma^x_i \sigma^x_j+J^{xz} \sigma^x_i \sigma^z_j + J^{zz} \sigma^z_i \sigma^z_j + J^{yy} \sigma^y_i \sigma^y_j.
\end{equation*}
It is similar to the popular Hamiltonian proposed in the context of quantum spin ice in Ref. S\!\!\citenum{gingras1},S\!\!\citenum{rossprx}, except that it does not contain the bond dependent terms $\gamma_{ij}$ and $\xi_{ij}$. With these results in hand, we can make the connection with the model of Eq. (\ref{hh}):
\begin{equation*}
{\cal H}_{1/2}= + \frac{1}{2}\sum_{i,j} 4\mu^2 {\cal J}^{zz}_{i,j} \sigma^z_{i}  \sigma^z_{j}+ \frac{1}{2}\sum_{i,j}  4v_y^2 \kappa_{i,j} \sigma^y_{i}  \sigma^y_{j},
\end{equation*}
where $\kappa_{i,j}$ is the coupling constant between octopoles. The renormalised couplings of the pseudo spin 1/2 are then:
\begin{eqnarray*}
J^{zz}  & = & 4\mu^2 {\cal J}^{zz} = (g_z/g_J)^2 {\cal J}^{zz} = {\cal J}',\\
J^{yy}  & = & 4 v_y^2 \kappa = {\cal K},
\end {eqnarray*}
and we recover the form discussed in the main text :
\begin{equation}
{\cal H}_{1/2}= + \frac{1}{2}\sum_{i,j} {\cal J}' \sigma^z_{i}  \sigma^z_{j}+ \frac{1}{2}\sum_{i,j}  {\cal K} \sigma^y_{i}  \sigma^y_{j}.
\label{h12}
\end{equation}

\begin{table}
\begin{tabularx}{\linewidth}{*{8}{>{\centering\arraybackslash}X}}
\hline \hline
$B_{20}$& $B_{40}$ &$ B_{43}$ & $B_{60}$ &$ B_{63}$ & $B_{66}$ & $g_{\parallel} $& $g_{\perp} $ \\ \hline
-190&5910&110&2810&10&-890&4.5&0 \\ \hline \hline
&&&&&&&
\end{tabularx}
\begin{tabularx}{\linewidth}{p{0.6cm}*{8}{c}>{\centering\arraybackslash}X} \hline \hline
 $J_z$& $-9/2$ & $-7/2$ & $-5/2$ & $-3/2$ & $\pm1/2$ & $3/2$ &$ 5/2$ &$ 7/2$ & $9/2$ \\ \hline
& $a$ & &  & $b$ &  & $c$ &  &  & $d$ \\ \hline
$| \uparrow\rangle$   & -0.878 & 0 & 0 & $-0.05$ & 0 & 0.476 & 0 & 0 & $0.009$ \\
$| \downarrow\rangle$ &-0.009 & 0 & 0 & 0.476 & 0 & 0.05 & 0 & 0 & -0.878 \\
\hline \hline
\end{tabularx}
\caption{Wybourne coefficients (in K), $g$-factors and ground state wave functions of \nd\ in \ndzr\, reproduced from Ref. S\!\!\citenum{Lhotel15b}. }
\label{DO}
\end{table}

\section{Mean field approximation}

The mean field Hamiltonian derived from Eq. (\ref{hh}) is given by: 
\begin{eqnarray*}
{\cal H}_{\rm MF} &=& \sum_i {\cal H}_i, \\
{\cal H}_i &=&
{\cal H}_{\mbox{CEF}} + \vec{J}_i \cdot\sum_{j} {\cal J}_{i,j} \cdot \langle \vec{J}_j \rangle + {\cal O}_i \cdot \sum_{j} {\cal K}_{i,j} \cdot \langle{\cal O}_j\rangle.
\end{eqnarray*}
Starting from a random configuration for the various observables, the contribution to ${\cal H}_{\rm MF}$ at site $i$ (in the unit cell) is diagonalised in the Hilbert space defined by the $\left\{ | J_z \rangle \right\}, J_z=-J,...,J$ basis vectors ($2J+1$ vectors). This yields the energies $E_{i,n}$ and the wave functions $\vert \phi_{i,n} \rangle$. The updated expectation values, $\langle A_i  \rangle'$, at each step of the iteration procedure, are given by: 
\begin{equation*}
\langle A_i  \rangle' =  \sum_{n} \frac{e^{-E_{i,n}/k_B T}}{Z} \langle  \phi_{i,n}  | A_i | \phi_{i,n} \rangle,
\end{equation*}
with
\begin{equation*}
Z =\sum_{n} \exp\left({-E_{i,n}/k_B T}\right).
\end{equation*}
These are then used to proceed to site $j$, and this is repeated until convergence. This procedure can be applied for both models given by both Eq. (\ref{hh}) and Eq. (\ref{h12}).

Figure S\ref{fig3-supmat}(a) shows the mean field phase diagram calculated at zero temperature as a function of ${\cal J'}$ and ${\cal K}$ for the model of Eq. (\ref{h12}). It contains an antiferromagnetic ``all in - all out'' phase (AIAO), a ferromagnetic ``2-in 2-out'' ordered  spin ice phase (SI); the ``O" prefix denotes non magnetic octopolar phases, either with an ``all in - all out'' nature, the octopolar moments on one tetrahedron being of the same sign (O-AIAO), or of the ``SI" nature, 2 octopolar moments being positive, the two others being negative (O-SI). 

The equations of the borders between the mean field phases can be obtained analytically based on the expressions of the different molecular fields: in the O-AIAO, AIAO, O-SI and SI, they are given by $-6{\cal K} \langle \sigma_y \rangle$, $-6{\cal J}' \langle \sigma_z \rangle$, $2{\cal K} \langle \sigma_y \rangle$ and $2{\cal J}' \langle \sigma_z \rangle$ respectively. As a result, it is possible to find the analytical expressions for the borders between the different phases:\\
i) between the O-AIAO and AIAO : ${\cal J}'={\cal K} $; \\
ii) between AIAO and O-SI, ${\cal J}'={-\cal K}/3 $; \\
iii) between O-SI and SI, ${\cal J}'={\cal K}$ \\
iv) and finally between SI and O-AIAO: ${\cal J}'=-3{\cal K}$. \\
The calculated temperature dependence of the octopolar moment for ${\cal J'}=1.2$~K and ${\cal K}=-0.55$~K is shown in Figure S\ref{fig3-supmat}(b). 

\section{Spin dynamics in the RPA approximation}

Spin dynamics are calculated in the random phase approximation (RPA). The excitations are constructed on the basis of the transitions between the mean field states $|\phi_{i,n} \rangle$ of energy $E_{i,n}$. In this spirit, the mean field Hamiltonian is written :
\begin{eqnarray*}
{\cal H}_{\rm MF} &= &\sum_{i, n} E_{i,n} | \phi_{i,n} \rangle \langle\phi_{i,n} | + {\cal V}, \\
{\cal V} &= & \sum_{i,j,n,m,n',m'} 
{\cal V}_{i,n,m,j,n',m'} | \phi_{i,n} \rangle \langle \phi_{i,m} | | \phi_{j,n'} \rangle \langle \phi_{j,m'} |,
\end{eqnarray*}
with
\begin{eqnarray*} 
{\cal V}_{i,n,m,j,n',m'} = \nonumber\\
\langle \phi_{i,n} |J_i - \langle J_i \rangle|\phi_{i,m} \rangle {\cal J}_{i,j} \langle \phi_{j,n'} |J_j -\langle J_j \rangle|\phi_{j,m'} \rangle \nonumber \\
 + \langle \phi_{i,n} |{\cal O}_i - \langle {\cal O}_i \rangle|\phi_{i,m} \rangle \kappa_{i,j}  \langle \phi_{j,n'}|{\cal O}_j -\langle {\cal O}_j \rangle|\phi_{j,m'} \rangle. \nonumber
\end{eqnarray*}
Following the generalised susceptibility approach $^{\rm S}$\cite{jensenb,kao03b,petit14b,robert15b}, the susceptibility reads:
\begin{eqnarray*}
\chi_{a,b}(Q,\omega) = \sum_{i,j,n,m,n',m'} \langle\phi_{i,n} |\mu^a - \langle \mu_i^a \rangle|\phi_{i,m} \rangle \times \\
\langle \phi_{j,n'} |\mu_j^b -\langle \mu_j^b \rangle|\phi_{j,m'} \rangle)~
e^{iQ.(\vec{r}_i-\vec{r}_j)}~R_{i,n,m | j,n',m'},
\end{eqnarray*}
where 
\begin{equation*}
\mu_i = \frac{g}{2} F(Q)~ J_i,
\end{equation*}
$g$ being the (possibly anisotropic) Land\'e factor, $F(Q)$ the magnetic form factor, and $R_{i,n,m | j,n',m'}$ is the solution of the matrix equation: 
\begin{equation*}
\left\{ (\omega + i \eta - \Delta ) I + \gamma L {\cal V} \right\} R = \gamma L.
\end{equation*}
$\eta$ is a convergence factor and 
\begin{eqnarray*}
L_{i,j,n,m,n',m'}   & = & \delta_{i,j} \delta_{m,n'} \delta_{n,m'} \\
\Delta_{i,n,m}  & = & \left(E_{i,n} - E_{i,m}\right)\delta_{i,j} \delta_{n,n'} \delta_{m,m'}\\
\gamma_{i,j,n,m,n',m'} & = & \left(p_{i,n} - p_{i,m}\right)\delta_{i,j} \delta_{n,n'} \delta_{m,m'}\\
p_{i,n} & = & \frac{\exp\left({-E_{i,n}/k_BT}\right)}{\sum_m \exp\left({-E_{i,m}/k_BT}\right) }.
\end{eqnarray*}
In this formalism, the dynamical structure factor consists of a series of well defined modes at energies $E_{Q,s}$ with spectral weight $I_{Q,s}$, where the $s$ index runs over $[(2J+1)^2 N_c]$ values, $N_c$ being the number of ions within the unit cell: 
\begin{eqnarray}
S(Q,\omega) = \frac{(2\pi)^3}{v_o}~(\gamma r_o)^2 \nonumber \\
\times \sum_{\tau,q,s} \delta(Q-\tau-q) \delta(\hbar \omega-E_{Q,s})~I_{q,s} \nonumber \\
+ \delta(Q-\tau+q) \delta(\hbar \omega+E_{Q,s})~I_{q,s}~~~~~~~~~
\label{sqwrpa}
\end{eqnarray}
where $\gamma$ is the gyromagnetic ratio, $r_o$ the classical radius of the electron,  $v_o$ the volume of the unit cell, and $\tau$ a reciprocal lattice vector.

The dynamical structure factor also encompasses an elastic contribution due to the ordered magnetic moments$^{\rm S}$\cite{lovesey}:
\begin{eqnarray*}
S_M(Q,\omega=0) = N~\frac{(2\pi)^3}{v_o}~(\gamma r_o)^2~ \sum_{\tau} \delta(Q-\tau)~\delta(\omega)  \\ 
\times \sum_{i,j} \sum_{ab} \langle \mu_i^a \rangle \left(\delta_{ab}-\frac{Q_aQ_b}{Q^2}\right) \langle \mu_j^b \rangle~e^{iQ.(\vec{r}_i-\vec{r}_j)},
\end{eqnarray*}
as well as a contribution due to the crystal structure$^{\rm S}$\cite{lovesey}: 
\begin{eqnarray*}
S_N(Q,\omega=0) = N \frac{(2\pi)^3}{v_o}~\sum_{\tau} \delta(Q-\tau)~\delta(\omega)  \\ 
\times \sum_{i,j} b_i b_j e^{iQ.(\vec{r}_i-\vec{r}_j)}~
\end{eqnarray*}
where $N$ is the number of unit cell, and the $b_i$'s are the scattering lengths of the various elements within a unit cell.

\section{Determination of the parameters}

The parameters ${\cal J'}$ and ${\cal K}$ of Hamiltonian (\ref{h12}) can be estimated by fitting the inelastic neutron scattering spectra. From calculations in the Random Phase Approximation (RPA), it is first observed that the existence of an inelastic signal can be accounted for only if the parameters are chosen in the octupolar phases (O-AIAO or O-SI). The physical reason is that in the magnetic phases, the ordered pseudo spin 1/2 is along the local $z$ ($\langle 111 \rangle$) axis. The transverse excitations thus involve spin components within the planes perpendicular to these axes. As those components have a Land\'e factor equal to zero, they cannot be observed with neutron scattering. Conversely, if the ordered component of the pseudo spin 1/2 is along the local $y$ axis (i.e. of octopolar nature in our case), the transverse excitations involve spin components along the local $z$ axis and thus become visible. 

\begin{figure}[h]
\includegraphics[width=9cm]{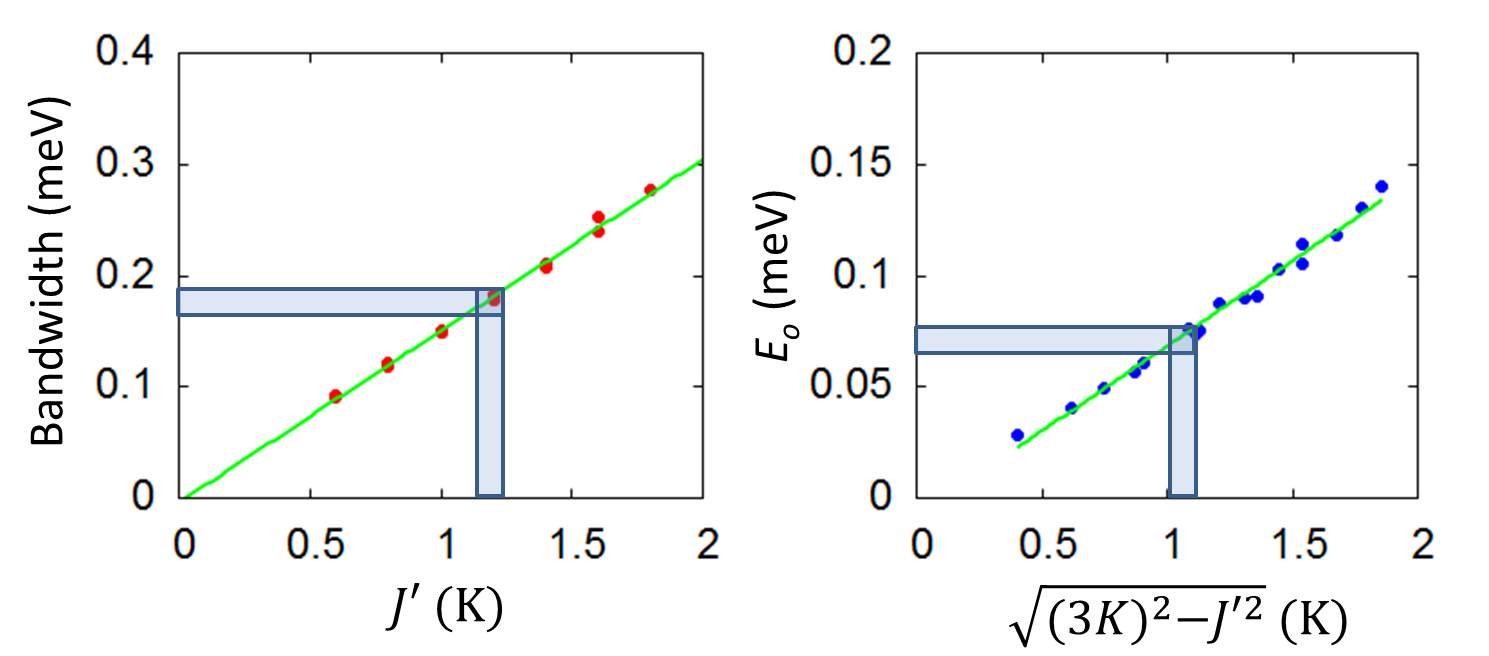}
\caption{\label{fig4-supmat} Evolution of the bandwidth (relative to $E_o$) and of $E_o$ as a function of ${\cal J'}$ and ${\cal K}$. Lines are a guide to the eyes. The blue horizontal and vertical bars show the uncertainties.}
\end{figure}

Calculations show empirically that the bandwith of the collective modes defined relatively to $E_o$ is proportional to ${\cal J'}$. It is also observed that the shift up to $E_o$ of the pinch point pattern is proportional to $\sqrt{(3{\cal K})^2-{\cal J'}^2}$, as illustrated in Figure S\ref{fig4-supmat}. The parameters along with their error bars can de obtained based upon this analysis. The best agreement is obtained for ${\cal J'} = (1.2\pm 0.025)$ K and ${\cal K'} = (-0.55\pm 0.025)$ K. 

\end{document}